\documentclass[aps,prl,twocolumn,amsmath,amssymb,longbibliography]{revtex4-2}
\usepackage{amsmath,amsfonts,amssymb,bm,hyperref,xcolor,amsopn}
\usepackage[capitalize]{cleveref}
\usepackage{graphicx,subfigure,epstopdf,multirow,diagbox}

\begin{document}
\title{Construction of a Pathway Map on a Complicated Energy Landscape}

\author{Jianyuan Yin$^{1}$}
\author{Yiwei Wang$^{2}$}
\author{Jeff Z.Y. Chen$^{3}$}
\email{jeffchen@uwaterloo.ca}
\author{Pingwen Zhang$^{1}$}
\email{pzhang@pku.edu.cn}
\author{Lei Zhang$^{4}$}
\email{zhangl@math.pku.edu.cn}

\affiliation{$^1$School of Mathematical Sciences, Laboratory of Mathematics and Applied Mathematics, Peking University, Beijing 100871, China.\\
$^2$Department of Applied Mathematics, Illinois Institute of Technology, Chicago, Illinois 60616, USA\\
$^3$Department of Physics and Astronomy, University of Waterloo, Waterloo, Ontario, Canada N2L 3G1\\
$^4$Beijing International Center for Mathematical Research, Center for Quantitative Biology, Peking University, Beijing, 100871, China.}

\date{\today}
\begin{abstract}
  { How do we search for the entire family tree of  possible intermediate states, without unwanted random guesses,
  starting from a stationary state on the energy landscape all the way down to energy minima?}
   {Here we introduce a general numerical method that constructs the pathway map,
   which guides our understanding of how a physical system moves on the energy landscape.}
  The method identifies the transition state between energy minima and the energy barrier associated with such a state.
  As an example, we solve the Landau--de~Gennes energy incorporating the Dirichlet boundary conditions to model a liquid crystal confined in a square box;
  we illustrate the basic concepts by examining the multiple stationary solutions and the connected pathway maps of the model.
\end{abstract}


\maketitle

\begin{figure}
  \centering
  \includegraphics[width=\columnwidth]{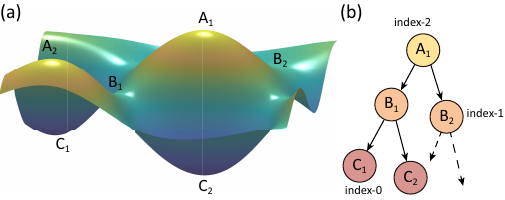}
  \caption{Illustration of
  (a) an energy landscape of a 2D energy surface and
  (b) the pathway map starting from the maximum, $A_1$ down to minima $C_1$ and $C_2$.
  Two saddle points $B_1$ and $B_2$ are connected to $A_1$.
  The indices labeled in (b) are those according to the Morse definition.
}
  \label{fig1}
\end{figure}

{\it {Introduction.}---}
A general mathematical-physics problem is minimization of an energy function depending on multiple variables. Its broad range of applications includes, but not limited to, the areas of condensed matter, soft matter, biophysics, materials sciences, financial physics, and beyond. The common theme is that the energy (free energy, or target function) landscape as a function of the physical variables in the system displays a multitude of minima  \cite{wales2003energy}.
The physical variables here could be the amino-acid locations in
the protein folding problem
\cite{bryngelson1995funnels, onuchic1997theory, leeson2000protein, laughlin2000middle, mallamace2016energy},
the atomic positions of an atomic cluster (such as the Lennard-Jones clusters) \cite{wales1997global, cameron2014computing},
the continuum field variables in an $AB$-diblock copolymer self assembly
\cite{fredrickson2002field, hamley2004developments, muller2005, FDRbook, cheng2010nucleation},
or even network parameters in a to-be-optimized artificial neural network \cite{goodfellow2016deep}.
One often makes a comparison of a two-variable (2D) problem with the geometric feature on a landscape: minima are basins, maxima are summits, and the two variables are the coordinates on a plane [see Fig.~\ref{fig1}(a)]; of great interest is the locations of minima that lie somewhere on the landscape.

{In this Letter, we tackle two long-standing, critical problems in computational physics: finding the global minimum and finding the relationship between the minima.} When a system has enough thermal excitations to overcome the energy barriers between energy minima, it undertakes a possible kinetic pathway on the energy landscape.
The minimal energy barrier that partitions two minima is associated with at least one   saddle point,
which is a transition state that the system likely experiences when it moves from one minimum to another \cite{asgeirsson2018exploring}.
The Morse index $m$ characterizes the nature of a saddle point  {in a $D$-dimensional problem}:
it is an energy maximum in $m$ directions, and minimum in the rest \cite{milnor2016morse}.
The $D$-variable energy surface possesses a complicated landscape;
the last two decades have witnessed progress in developing mathematical procedures to
determinate and understand saddle points numerically \cite{henkelman2002methods, vanden2010transition, zhang2016recent, zhang2016optimization}.

For example, in Fig.~\ref{fig1}(a), the {lowest} energy barrier between two minima $C_1$ and $C_2$ is located at a saddle $B_1$, not $B_2$.
Assessing the energy levels alone cannot determine the dynamic process.
The purpose of this Letter is to introduce the more informative concept of a pathway map through an analysis of the Hessian matrix, by asking the question of which eigendirection is unstable and what next saddle that unstable direction leads to.
For example, Fig.~\ref{fig1}(b) depicts how lower-index saddle points are connected to the higher-index ones and such a pathway map is more useful than simply connecting minima through index-1 saddles \cite{chen2015locating}.
The pathway map is established in a multidimensional configurational space and rules out the energetically close but configurationally far states.
In a high-$D$ system, the pathway map can be constructed from a parent state (a high-index saddle) all the way down to minima (index-0 {solutions}).
It is hence desirable, and indeed the purpose of this Letter, to design a numerically tractable algorithm that constructs a complete pathway map, which guides us on how the system runs into metastable states and the energetic requirement to overcome energy barriers when the system moves between the states.

Assume that the energy surface $E({\bf x})$ is a function of a $D$-dimensional variable ${\bf x}$.
All stationary points satisfy the nonlinear equations $\nabla E(\bf x)= 0$, and
the Morse index is determined by examining the Hessian $\nabla \nabla E(\bf x)$ at these points \cite{mehta2014potential}.
While these mathematical definitions are elementary, capturing all stationary points is practically difficult in computation.
Many algorithms, e.g., the minimax method \cite{li2001minimax, zhang2007morphology},
the deflation technique \cite{brow1971deflation, farrell2015deflation}, the eigenvector-following method \cite{wales1994rearrangements, doye1997surveying, doye2002saddle}, and the numerical polynomial homotopy continuation method \cite{mehta2011finding}
are dedicated to solving the nonlinear equations, usually relying on an initial guess that deterministically leads to a stationary point.
However, as more solutions are discovered, it becomes increasingly difficult to propose and fine-tune initial guesses, to search for remaining solutions.
A large number of optimization methods are also developed to directly find local minima;
it is often computationally expensive, as the total number of local minima usually grows exponentially in a high-$D$ problem
and a random search is not the best choice \cite{wales2003energy}.
{Therefore, from a computational perspective, it is essential to establish a numerical procedure to discover the pathway map \emph {systematically},
rooted from a high-index saddle as a parent and descending to minima.} The advantages of our method over other methods are discussed case by case in Supplemental Material \cite{SM}.

{\it {Downward Search.}---}
{Here we introduce a systematic computational method} that enables the search for such a map starting from an index-$m$ saddle. The assumption is that the location of the saddle point is at ${\bf x}^*$ and that the $m$ normalized vectors ${\bf v}_i^*$ corresponding to the $m$ negative eigenvalues of the Hessian matrix are known at this stage, where $i=1,2,3,...,m$.

The essential idea is to choose ${\bf x}(0) = {\bf x}^* \pm \epsilon {\bf u}$ as the initial search position for a lower index-$k$ ($k<m$) saddle with a small $\epsilon$ that ``pushes'' the system away from the index-$m$ saddle.
The direction of pushing, ${\bf u}$, is along a linear combination of $(m-k)$ vectors (whose negative eigenvalues have the smallest magnitudes) selected from
the set of $\{{\bf v}^*\}$.
The other $k$ orthonormal vectors from the set are used as the initial unstable directions for the next index-$k$ saddle.
This way, the system can gently roll off the original index-$m$ saddle point in the unstable direction within a controlled procedure.
Normally, a pair of index-$k$ saddles can be found this way, corresponding to the $\pm$ sign in the initial guess.

The details of this numerical procedure, the so-called $k$-HiOSD method \cite{yin2019high}, is documented in the Supplemental Material \cite{SM}, step-by-step.
It computationally determines an index-$k$ saddle point based on ${\bf x}(0)$ and ${\bf v}_i(0)$ ($i=1,2,3,...,k$) as an initial input.
By relaxing a pseudo-Langevin dynamics for time-dependent ${\bf x}$ and ${\bf v}_i$, the computation eventually arrives at all conditions that need to be satisfied by an index-$k$ saddle, producing ${\bf x}^*$ and ${\bf v}_i^*$ for it.

Hence, by repeating the above procedure starting from an existing index-$m$ saddle as a parent,
we establish a systematic scheme to search for {\emph {all}} saddle points  branched from this parent and to construct a family tree that eventually terminates at index-$0$ {solutions (minima)}, the local energy minima.
This avoids the pitfalls existing in previous numerical schemes to search for an arbitrary solution starting from random initial guesses with no control on both the search direction and Morse index of the final numerical outcome.
The family tree provides the crucial information on the relationships between saddle points.
In particular, the pathway between two lower-index saddle points is clearly mapped out through a higher-index saddle point.

\begin{figure*}
  \includegraphics[width=\linewidth]{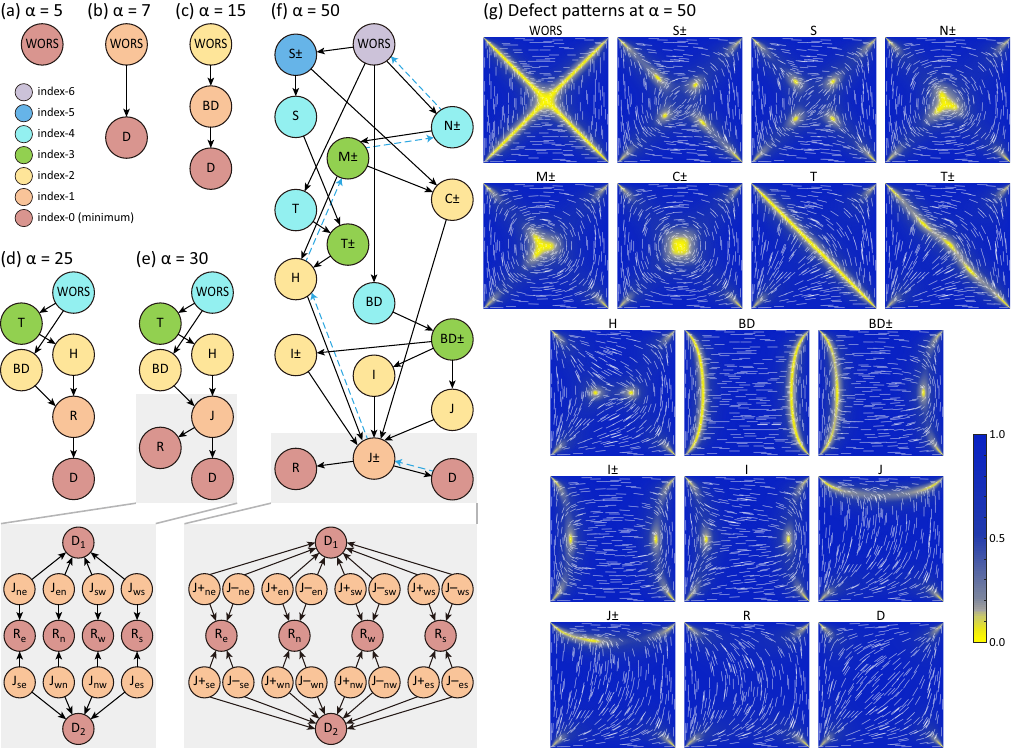}
  \caption{Pathway maps found from a confined 2D LdG model with parallel orientational alignment at the boundaries.
  From (a) to (f), the evolution of the pathway maps are illustrated as $\alpha$, a system parameter proportional to the area of the confining square, increases from (a) $\alpha=5$, (b) $7$, (c) $15$, (d) $25$, (e) $30$, to (f) $50$.
  The color of the nodes specifies the Morse index of saddle points.
  The solid arrows indicate how lower-index saddles can be deduced from higher ones.
  An example of upward search is represented by the dashed arrows.
  The height of a node approximately corresponds to its energy.
  The seventeen defect states are illustrated on the right panel, where the color represents the relative magnitude of directional ordering and the white bars the nematic field directions \cite{SM}.
  The insets further explain the map details discussed in the text.}
  \label{LdGpath}
\end{figure*}

{\it {Liquid crystals in 2D confinement}---}
To {demonstrate the success of the method}, we present here the pathway map of 2D nematic liquid crystals confined in a square box (see illustrations in Fig.~\ref{LdGpath}).
The mathematical model we use is the Landau--de~Gennes (LdG) free-energy
in reduced units \cite{de1995physics}
\begin{equation}\label{LdGdef}
E[{\sf Q}({\bf r})]
  =\int d {\bf r} \left\{ \frac{1}{2} {\big|\nabla {\sf Q} \big|^2}
   +\alpha f_{b}[{\sf Q}({\bf r})] \right\},
\end{equation}
where a $2\times 2$ traceless and symmetric tensor field ${\sf Q}({\bf r})$, characterizing the orientational ordering of the liquid crystals, is considered.
Only two elements of the ${\sf Q}$-tensor are independent.
The variable ${\bf r}=(x,y)$ resides in the domains $[-1,1]$ for $x$ and $[-1,1]$ for $y$.
The specifics of the LdG model, including the Landau function $f_{b}$ and
Dirichlet conditions that require the alignment of the liquid crystal directors at the four  boundaries, are given in the Supplemental Material \cite{SM}.
The system parameter $\alpha$ is directly linked to the area of the confining square.

The energy functional in Eq.~\eqref{LdGdef} must be minimized with respect to the tensor field ${\sf Q}({\bf r})$ for a given $\alpha$.
There are two equivalent approaches to finding a stationary solution.
The first is to take and solve the Euler-Lagrange equation associated with the unknown field ${\sf Q}({\bf r})$ \cite{robinson2017molecular, wang2019order};
the second, closer to the spirit presented here, is to discretize the $(x,y)$ region by a grid system containing $G$ grid points and to treat the two independent elements of $\sf Q$ at these points as individual variables;
the stationary solution of the energy in Eq.~\eqref{LdGdef} is then obtained simply by equating the gradient of $E$ in the $D$-dimensional space to zero,
where $D=2G$ are the total number of individual variables \cite{davis1998finite, wang2017topological}.

When $\alpha$ is sufficiently small, it was shown that the so-called well order-reconstruction solution (WORS) is the only stationary solution and corresponds to the global minimum of Eq.~\eqref{LdGdef} \cite{canevari2017order}.
Displayed in Fig.~\ref{LdGpath}(a) is such a state with index-0 (minimum).
WORS was suggested as a possible stationary state based on the Onsager model in Ref.~\cite{chen2013structure} and the LdG model in Ref.~\cite{kralj2014order}.

\begin{table}[!b]
  \renewcommand\arraystretch{1}
  \centering
  \begin{tabular}{c|cccccccccccc}
    \hline\hline
    $\alpha$    & 5 & 7 & 15 & 25 & 35 & 50 & 75 & 90 & 100 & 130 & 160 & 200 \\ \hline
    Morse index & 0 & 1 & 2  & 4  & 5  & 6  & 8  & 9  & 10  & 12  & 13  & 14  \\
    \hline\hline
  \end{tabular}
  \caption{Morse index of WORS at different $\alpha$ based on the LdG model.}\label{WORS}
\end{table}

One can show that WORS remains a stationary point for all $\alpha$ \cite{canevari2017order}, but the Morse index of WORS changes \cite{wang2019order}.
Numerically, by using WORS at a smaller $\alpha$ as an initial guess, we can always obtain WORS at a larger $\alpha$.
Table \ref{WORS} summarizes the Morse index of WORS as a saddle point at different $\alpha$.
Physically, as $\alpha$ increases, the larger confining area accommodates the development of other more complicated defect states.
The parent WORS is no longer a stable state and the system begins to descend to other stable states.

Using our $k$-HiOSD method, we can exhaustively enumerate all possible states starting from WORS.
Figure \ref{LdGpath} presents the pathway maps for a number of typical $\alpha$, where the defect states specified in the nodes are illustrated on the right panel.
Take Fig.~\ref{LdGpath}(c), for example.
At $\alpha=15$, WORS is an index-2 saddle,
which bifurcates to two boundary distortion (BD) states (related by $\pi/2$ coordinate rotation) that are index-1 saddles.
Both BD states then further relax to two diagonal ($D$) states (again related by $\pi/2$ coordinate rotation) as the global minima.
This can be compared to the pathway map in 2(b) when $\alpha=7$, where the index-1 WORS is the direct pattern from which the two $D$ states bifurcates.
Assume that we wish to design a physical device to take advantage of the different optical properties of the dual $D$ states by switching between them \cite{tsakonas2007multistable};
the energy barrier that the device has to overcome and the intermediate states between the two $D$ states are different for system size $\alpha=7$ (b) and $\alpha=15$ (c);
the former goes through WORS and the latter through BD.

As $\alpha$ increases further, the pathway map become more complicated.
At $\alpha=30$, WORS is now an index-4 saddle.
Two types of unrelated defect states, $D$ and rotated ($R$) (and their related counterparts by $\pi/2$ rotation), emerge as index-0 minima.
A different lesson from the above is learned;
the dual $D$ states are connected by a metastable $R$ state, following the dynamic pathway sequence $D_1\rightarrow J_{\mathrm{ne}}\rightarrow R_{e}\rightarrow J_{\mathrm{se}}\rightarrow D_2$, or vice versa, where the subscripts indicate the rotated orientations of the pattern \cite{kusumaatmaja2015free}.
The system could be trapped in $R$ as a metastable state because it is an energy minimum.
In order to move to $D$ where the global minimum resides, the $R$ state needs to overcome an energy barrier $E_{J} - E_{R}$, which is available from our calculation.
From the Kramers theory \cite{kramers1940brownian}, we can then estimate the trapping time to find out the relaxation kinetics of the bistable system \cite{bhattacharjee1989kramers}.

The relationships between the stationary states could become quite complicated at $\alpha=50$, as Fig.~\ref{LdGpath}(f) demonstrates.
WORS is now index-6.
Starting from WORS as the parent, we search for other states generation by generation to produce the entire family tree
(see an illustration of the dynamic downward pathway sequence: $\mathrm{WORS} \rightarrow \mathrm{BD} \rightarrow \mathrm{BD}+\rightarrow I\rightarrow J-\rightarrow D$ in Supplemental Material \cite{SM}).
In total, 89 distinct solutions can be found and after classification to account for basic symmetry operations, they fit into 17 basic types.
Surprisingly, the structure that contains more defect features, $J{\pm}$, undertakes the simpler $J$ as the transition state that partitions a $D$ state (global minimum) and an $R$ state (local minimum).
The two ground states $D_1$ and $D_2$ are connected by pathway sequences $D\rightleftarrows J\pm \rightleftarrows R \rightleftarrows J\pm\rightleftarrows D$
as illustrated in the inset where the orientations of the involved states decide the exact pathway.

One of the technical advantages of our method is to produce the entire family tree under a parent.
A direct result is our finding of states $N\pm$, $M\pm$, $S\pm$, and $T\pm$, which were not reported in the previous numerical studies of the same LdG model \cite{robinson2017molecular} or the Onsager model \cite{yao2018square}.
Previously, the variety of defect states of these models were obtained by simply solving the necessary condition that the gradient of the energy must be zero, without the systematic inquiry of the sufficient condition of whether the found states are high-index saddle points.
The conclusion we draw for the LdG model, that WORS is stable at small $\alpha$ and $D$ is stable at large $\alpha$ with $R$ as the possible metastable state, is consistent with the phase diagram concluded in Ref.~\cite{yao2018square} based on the Onsager model.
All these can be further compared with recent experimental observations where no patterns having high Morse index were actually stabilized  \cite{galanis2006spontaneous, tsakonas2007multistable, galanis2010nematic, silva2011self, lewis2014colloidal}.

{\it {Upward Search.}---}
The downward search strategy guarantees the systematic finding of energy minima as index-0 solutions starting from a given parent,
which is already superior to most other optimization numerical methods.
On the other hand, if multiple parent states exist, we need to conduct the search on each family tree starting from an individual parent state.
For example, starting from $A_2$ in Fig.~\ref{fig1}(a), a family tree different from the one in Fig.~\ref{fig1}(b) would be found.
The $A_1$ and $A_2$ trees may share a common $B_1$ node but the downward searches establish separated maps.

The key to find complete, interwound pathway maps is to conduct a systematic upward search for the parent states.
The $k$-HiOSD algorithm already embeds a mechanism that allows us to search upward starting from an index-$m$ saddle point to find an index-$k$ saddle where $k>m$.
Again, we assume that all eigenvalues and eigenvectors of the Hessian matrix are known.
To do so, the initial condition ${\bf x}(0) = {\bf x}^* \pm \epsilon {\bf u}$ perturbs ${\bf x}^*$ along the direction $\bf u$, which is selected along a linearly combination of $(k-m)$ eigenvectors corresponding to
small positive eigenvalues.
The initial $\{{\bf v}^*\}$ set used in the search includes these $(k-m)$ eigenvectors and the original $m$ eigenvectors of the negative eigenvalues.
By doing so, the search may branch into other family trees through, e.g., the shared $B_1$ in Fig.~\ref{fig1}(a) to end up at parent $A_2$.
Then a downward search from $A_2$ covers the entire pathway map different from the one illustrated in Fig.~\ref{fig1}(b).

Take the scenario where $D$ is a known, initial state without the knowledge of existence of other structures in Fig.~\ref{LdGpath}(f).
Along the upward pathway indicated by dashed arrows, in a single search we can produce the upward pathway sequence $D \rightarrow J+\rightarrow H\rightarrow M+\rightarrow N+\rightarrow \mathrm{WORS}$ (see Supplemental Material \cite{SM}).
Once WORS is arrived at, no other upward searches give rise to a solution with a higher index, hence WORS is found as a parent.

{\it{Summary.}---}
This Letter introduces the concept of pathway maps starting from parent states, illustrating
the relationships between the stationary states.
Our $k$-HiOSD numerical algorithm introduced here is designed for this purpose.
As an example, we determine the solutions of the LdG model for confined liquid crystals and discuss the usefulness of such a pathway
map in understanding the physical properties of a multisolution problem, which are otherwise unobtainable by
other methods.

There are at least three important aspects of this approach.
The pathway map starts with a parent structure and then relates the entire family completely down to energy minima.
The first is the tight control of the initial conditions, which overcomes the difficulty of tuning initial guesses to search stationary points needed in other existing methods.
The second is our emphasis on the relationships between \emph{all} stationary states, established by the pathway map.
Most importantly, such connections reveal the hidden physical processes and the transition states are clearly shown on the map, which guide our understanding of a physical system.
The concept of pathway maps has not been suggested in previous approaches on similar problems.
The third is the capability of upward search with a selected direction so that the entire search navigates up and down on the energy landscape, as long as the saddle points are connected somewhere.
The procedure offers a general mechanism for the complete determination of all energy minima (hence the global energy minimum) without limitations on energy types, if the number of stationary states is finite.
This opens the door to find the dynamic pathways on a complicated energy landscape, which is a critical enabler for many mathematical problems in physics and engineering.

We thank Professor Yucheng Hu for helpful discussions.
This work was supported by the National Natural Science Foundation of China (Grants No.~11622102, No.~11861130351, No.~21790340, No.~11421101 and No.~21873009).
J.~Y. acknowledges the support from the Elite Program of Computational and Applied Mathematics for Ph.D. Candidates of Peking University.

\bibliography{WORSbib}

\end{document}


\title
{Supplemental Material:\\Construction of a Pathway Map on a Complicated Energy Landscape}

\author{Jianyuan Yin$^{1}$}
\author{Yiwei Wang$^{2}$}
\author{Jeff Z.Y. Chen$^{3}$}
\author{Pingwen Zhang$^{1}$}
\author{Lei Zhang$^{4}$}

\affiliation{$^1$School of Mathematical Sciences, Laboratory of Mathematics and Applied Mathematics, Peking University, Beijing 100871, China.\\
$^2$Department of Applied Mathematics, Illinois Institute of Technology, Chicago, Illinois 60616, USA\\
$^3$Department of Physics and Astronomy, University of Waterloo, Waterloo, Ontario, Canada N2L 3G1\\
$^4$Beijing International Center for Mathematical Research, Center for Quantitative Biology, Peking University, Beijing, 100871, China.}

\date{\today}
\maketitle

\section{$k$-HiOSD method}\label{AAA}

Here, we explain the essential steps in the high-index optimization-based shrinking dimer method
for an index-$k$ saddle point ($k$-HiOSD method) to find an index-$k$ saddle point based on an initial guess \cite{yin2019high}.
The energy surface $E=E({\bf x})$ is assumed at least twice differentiable and $\bf x$ is a $D$-dimensional vector variable.

An index-$k$ saddle point at ${\bf x}^{*}$ has the following properties.
It is a local maximum in $k$ {orthonormal} directions and {minimum} in $(D-k)$ {orthonormal} directions in the $D$-dimensional space.
These can be understood from the Hessian in a matrix form with the elements
\begin{equation}\label{Hdef}
H_{ij} = {\partial \over \partial x_i} {\partial \over \partial x_j} E({ x_1, x_2, ... x_D}) \Big |_{{\bf x} ={\bf x}^*}
\end{equation}
where $i,j = 1,2,3,..., D$, or in the tensor form,
\begin{equation}\label{Htensor}
{\sf H} = \nabla\nabla E({\bf x}) \Big |_{{\bf x} ={\bf x}^*}.
\end{equation}
It has $D$ eigenvalues: $(D-k)$ positive and $k$ negative.
Let ${\bf v}_i^{*}$ be the normalized eigenvector corresponding to the $i$th smallest eigenvalue, where $i=1,..,k$.

The $k$-HiOSD method locates such a saddle point by solving the following dynamic problem from an initial guess,
where both ${\bf x}$ and ${\bf v}_i$ are treated as functions of time $t$ so that their time derivatives $\dot {\bf x}$ and $\dot {\bf v}_i$ are considered.
The pseudodynamics that enables the numerical search for an index-$k$ saddle point is assumed to satisfy
\begin{equation}\label{hiosd}
\left\{
\begin{aligned}
 \dot {\bf x}   &=- \beta\left({\sf{I}}-
 2 \sum_{i=1}^k {\bf v}_i {\bf v}_i\right)\cdot \nabla E({\bf x}),\\
 \dot {\bf v}_i  &=-\gamma\Bigg({\sf{I}}-{\bf v}_i{\bf v}_i-2
 \sum_{j=1}^{i-1}{\bf v}_j{\bf v}_j\Bigg)\cdot {\bf{h}}({\bf x}, {\bf v}_i),
\end{aligned}
\right.
\end{equation}
with the orthonormal constraints $ {\bf v}_i\cdot {\bf v}_j=\delta_{ij}$.
Note that ${\sf I}$ is a $D$-rank unit tensor.
A line derivative is numerically taken on the vector field $\nabla E({\bf x})$ to represent the projection of the Hessian in direction $\bf v$,
\begin{equation}\label{lineHdef}
{\bf {h}}({\bf x}, {\bf v})  = \left[
\nabla E({\bf x}+ h {\bf v}) -
\nabla E({\bf x}-h {\bf v})
\right]/2h,
\end{equation}
where $h$ is a small number.
To produce the example in the text, we used $h=10^{-6}$.
The relaxation parameters,  $\beta >0$ and $\gamma >0$, guide the rate of the dynamics and are usually taken to be small parameters, according to the actual problem.
For example, in the text, when we deal with the liquid-crystal confinement problem, we let
$\beta= 5\times 10^{-5}$ and $\gamma=10^{-5}$.
To complete the differential equations, the initial conditions (initial guess), ${\bf x}(t=0), {\bf v}_i(t=0)$, are required.

The time evolution of the involved quantities can be numerically treated by discretization the $t$ variable of Eq.~\eqref{hiosd},
to be used in the explicit Euler scheme, the Barzilai--Borwein gradient method \cite{barzilai1988two}, or any other numerical schemes.
One can show that once the $k$-HiOSD dynamics stably converges to a final solution ${\bf x}^*$  and ${\bf v}_i^*$,
i.e., when the left hand sides of the equation set completely vanish to zero, all
properties of an index-$k$ saddle point are satisfied \cite{yin2019high}.
Note that no Hessian matrix and no eigenvalue calculation of it is explicitly needed in this method.
The algorithm requires an initial guess on ${\bf x}(t=0)$ and ${\bf v}_i(t=0)$ as the input and yields $\bf x^*$ and ${\bf v}_i^*$ as the output.
A converged set ${\bf v}_i^*$ guarantees that their eigenvalues to be negative.

On the other hand, if the above pseudodynamics does not converge, then an index-$k$ saddle point does not exist starting from the initial search position ${\bf x}(t=0)$.
In this case, either a new guess is made or another $k$ is attempted.

\section{Landau--de~Gennes model for confined liquid crystals in 2D}\label{NLC}
One of the best known models to describe the liquid-crystal ordering is the Landau--de~Gennes (LdG) model, which treats the free energy as a functional of the ${\sf Q}$-tensor field \cite{de1995physics,mottram2014introduction}.
Minimizing the free energy, one finds the stable liquid-crystal configurations of the system.
In the two-dimensional example used in the text, ${\sf Q}$ is taken to be a $2\times 2$ traceless and symmetric tensor \cite{kralj2011curvature, robinson2017molecular},
\begin{equation}\label{Qdef}
{\sf Q} =\begin{bmatrix}
Q_{11}({\bf r}) & Q_{12}({\bf r}) \\
Q_{12}({\bf r}) & -Q_{11}({\bf r})
\end{bmatrix}
\end{equation}
with scalar components $Q_{11}({\bf r})$ and $ Q_{12}({\bf r})$.
We use reduced Cartesian coordinates $x, y$ for the two components of ${\bf r}$, where $x$ and $y$ are both variables in the domain $[-1,1]$ for square confinement.

The particular form of the LdG model used here is written in reduced units,
\begin{equation}\label{LdG}
E[{\sf Q}({\bf r})]
  =\int d {\bf r} \left\{   \frac 1 2 {\big|\nabla {\sf Q} \big|^2}
   + \alpha f_{b} [{\sf Q}({\bf r})] \right\}
\end{equation}
where $|\nabla {\sf Q}|^2 \equiv \sum_{ij}[({\partial Q_{ij} / \partial x})^2+({\partial Q_{ij} / \partial y})^2]$
and $\alpha$ is a positive system parameter that is inversely proportional to the bending modula under the one-elastic-constant assumption and proportional to the overall area of the confined system \cite{de1995physics}.
The bulk free energy in reduced units is
\begin{equation}\label{eqn:nlcenergy2d}
f_{b} [{\sf Q}({\bf r})] =
\frac{a}{4} \big| {\sf Q} \big|^2 +\frac{1}{8} \big| {\sf Q} \big|^4,
\end{equation}
where $|{\sf Q}|^2=\mathrm{tr}({\sf Q}^2)$, and $a$ is the reduced temperature difference.
To ensure that the system is in the nematic state, we choose a low ``temperature'' and let $a=-1.672$.

In a uniform bulk phase where there is no ${\bf r}$-dependence, in a diagonalized form the $\sf Q$-tensor can be written as
\begin{equation}\label{S1def}
{\sf Q} = {1\over 2}\begin{bmatrix}
S & 0 \\
0 & -S
\end{bmatrix},
\end{equation}
where $S$ is the scalar orientational order parameter.
Minimizing $f_{b}$ with respect to $S$ we have $S = \pm S_0$ where $S_0 = {\sqrt{2|a|}}$.
Returning to the spatially inhomogeneous case, we force the four Dirichlet boundary conditions to follow
\begin{equation}\label{S2def}
{\sf Q}(x=\pm 1, y  )= {S_0\over 2}\begin{bmatrix}
1 & 0 \\
0 & -1
\end{bmatrix},
\end{equation}
and
\begin{equation}
{\sf Q}(x, y=\pm 1  )= {S_0\over 2}\begin{bmatrix}
-1 & 0 \\
0 & 1
\end{bmatrix}.
\end{equation}
That is, the ${\sf Q} ({\bf r})$ tensor at the boundaries simply adopts the bulk value but with the nematic director aligning in parallel to the square boundary lines.

After a solution is found from solving the pseudodynamics according to the $k$-HiOSD method, the generally nonuniform, undiagonalized ${\sf Q} ({\bf r})$ is analyzed by calculating its eigenvalues,
\begin{equation}\label{lambda}
S ({\bf r})/2=  \pm {\sqrt {Q_{11}^2({\bf r})+Q_{12}^2({\bf r})}}.
\end{equation}
The color in Fig.~2 of the text represents the magnitude of $|S ({\bf r})|/S_0$.
A nematic region has $|S|/S_0=1$ and isotropic state $|S|/S_0=0$.
Also plotted in Fig.~2 are white lines that represent the nematic field director, which is deduced from the vector
\begin{equation}\label{ndef}
{\bf n}({\bf r}) = {\hat x}\cos \theta ({\bf r}) + {\hat y}\sin \theta ({\bf r})
\end{equation}
with
\begin{equation}\label{thetadef}
\cos \theta({\bf r})  = \sqrt{1/2 + Q_{11}({\bf r})/|S({\bf r})|}.
\end{equation}
No arrows are drawn on these vectors. The unit vectors along the $x$- and $y$-axes are ${\hat x}$ and $\hat y$.

\begin{figure*}
  \centering
  \includegraphics[width=12.7cm]{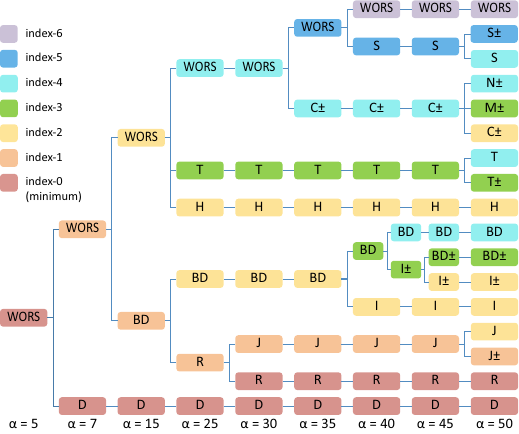}
  \caption{The bifurcation diagram of a confined 2D LdG model.
  Each node represents a type of defect states as specified in Fig.~2 of the text.
  The color of the nodes specifies the Morse index of saddle points.
  Each line represents a solution branch, and a T-junction represents a pitchfork bifurcation.
  }
  \label{fig3}
\end{figure*}

At different $\alpha$,
displayed in Fig.~2 of the text, multiple defect states are sequentially found by constructing the pathway maps.
New defect solutions can be well explained through the solution bifurcation diagram in Fig.~\ref{fig3} here. Taking the simple case $\alpha=7$ as an example, WORS bifurcates into two $D$ states via a pitchfork bifurcation and becomes an index-1 saddle.
Then this index-1 saddle WORS bifurcates into two BD states via another pitchfork bifurcation and becomes an index-2 saddle itself when $\alpha$ increases to 15.
As a result, two $D$ states (minima), two BD states (index-1 saddles), and WORS (an index-2 saddle) exist at $\alpha=15$.
The emergence of $R$ states as minima can also be explained from this diagram.
The $R$ states are bifurcated from the BD states and stabilized via a pitchfork bifurcation from index-1 saddles to minima.
Another complicated bifurcation is the simultaneous emergence of $T$ and $H$ states resulted from a pair of repeated eigenvalues of WORS, and similar processes occur when $N\pm$ and $M\pm$ states are bifurcated from $C\pm$ states.
After multiple bifurcations, finally all 17 types of defect states at $\alpha=50$ are produced, as shown in Fig.~\ref{fig3}.

\begin{figure}
  \centering
  \includegraphics[width=\linewidth]{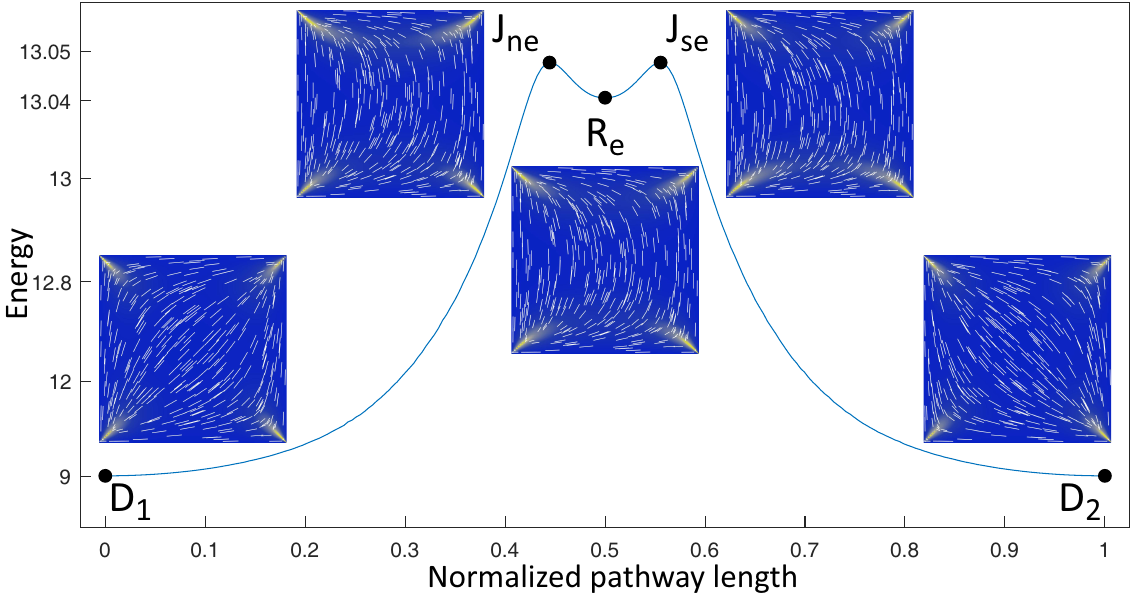}
  \caption{The pathway sequence $D_1\rightarrow J_{\mathrm{ne}}\rightarrow R_{e}\rightarrow J_{\mathrm{se}}\rightarrow D_2$ between two dual $D$ states at $\alpha=30$.
  The vertical axis is stretched according to $-\log(13.06-E)$ for better visualization, and the horizontal axis describes the normalized pathway length from the $D_1$ state.}
  \label{fig4}
\end{figure}

The pathway map provides the relationship between the stationary states, in addition to the information on the energy levels.
For example, at $\alpha=30$, the pathway map [see Fig.~2(e) of the text] shows that the dual $D$ states are connected by a metastable $R$ state via the dynamic pathway sequence
$D_1\rightarrow J_{\mathrm{ne}}\rightarrow R_{e}\rightarrow J_{\mathrm{se}}\rightarrow D_2$.
This connection can only be realized in the multi-$D$ space, by an analysis of eigendirections of the Hessian, which connects the configurations, rather than energy levels.
Figure \ref{fig4} here further demonstrates how the configurational patterns changes through this connection.

Lower Morse-index states are usually symmetry-breaking states connected to the higher one in the pathway map.
For example, at $\alpha=50$, the unstable $N\pm$ and $M\pm$ states could be misidentified to contain a 3-fold symmetry near the defect centers.
They actually represent a contraction of the four-fold symmetry in WORS, with one diagonal ``arm'' diminished to form a 3-branched center defect (no symmetry here).
Hence they are symmetry-broken, transit states from WORS (4-fold) to lower-index states.
A careful examination reveals that for the illustrated states in Fig.~2(g) of the text, the three arms of the $N+$ defect center consist of three $+1/2$ defects at angles $135^\circ, 225^\circ, 315^\circ$ from the horizontal line and a $-1/2$ defect at the center.
The three arms of the $M+$ defect center consist of three $+1/2$ defects at angles $=0^\circ, 135^\circ, 225^\circ$ and a $-1/2$ defect at the center.
These angles are reminiscent from the four-fold symmetry of WORS.

\section{Advantages over other numerical methods}\label{Comparison}
Finding all stationary points is difficult in computation, practically.
Here, we report a novel computational method to construct a pathway map that not only consists of all stationary points but also shows the connections between them, which, to the best of our knowledge, is the first time in the existing literature.

In the research areas of minimization and rare events, most studies focus on developing efficient methods to find index-1 saddle points only, in contrast to the current work emphasizing on all saddles on the entire landscape, regardless of their indices. A popular approach is to use various versions of the surface-walking methods.
Typically, a search starts from a single state on the energy landscape and requires the first or second derivatives of the energy to locate index-1 saddle points without {\it a priori} knowledge on the existence of the final state.
Even if an index-1 saddle is found, its relationships with other index-1 saddles are uncertain.
Some representative methods include the eigenvector-following method \cite{cerjan1981finding,wales1994rearrangements, doye1997surveying}, the dimer-type methods \cite{henkelman1999dimer,zhang2012shrinking,zhang2016optimization}, the gentlest ascent dynamics \cite{weinan2011gentlest,gao2015iterative}, the activation-relaxation technique \cite{cances2009some, machado2011optimized}.
We refer to \cite{vanden2010transition,henkelman2002methods,zhang2016recent} for excellent reviews.

The eigenvector-following method, for example, searches the index-1 saddles by selectively maximizing the energy in one direction and simultaneously minimizing in all other directions \cite{wales1994rearrangements,doye1997surveying,doye2002saddle}.
The maximizing direction is obtained from finding the eigenvector corresponding to the smallest eigenvalue of the Hessian.
In another example, the optimization-based shrinking dimer (OSD) algorithm is also designed to find index-1 saddles \cite{zhang2016optimization}.
A dimer method makes use of first derivatives of the energy only via a dimer system, and works by alternately performing rotation and translation steps.
The OSD algorithm constructs the rotation and translation steps in the dimer method under an optimization framework, and then applies the Barzilai--Borwein gradient method as an effective implementation to compute the index-1 saddles.

To search for other index-1 and minimum stationary points, both the eigenvector-following and OSD algorithms need to start the search from a local minimum, choose a random eigenvector direction to follow, and find a connected (occasionally unconnected) index-1 saddle.
Once an index-1 saddle is found, one can find the minima that it connects to.
This searching procedure is repeated until no new index-1 saddles and minima can be found computationally.
They both suffer from the problems of redundant findings and can miss hard-to-find minima.

We treat the search using a different philosophy.
The downward search algorithm constructs the pathway map starting from a parent state (high-index saddle), down to form the entire family tree, avoiding the difficulty of tuning initial guesses in the above two methods.
It provides an efficient approach to find multiple minima systematically.
In the case of a large number of minima, any random guess work on the initial conditions, currently needed in most existing methods, would make unguided searches and hence prolong the search time.
We fix this major problem by guiding the search through saddle points, order by order, without tuning any random guesses.
Some minima (or even a hidden global minimum) are hard to reach simply because of the very narrow kinetic pathway, which makes the random-guess search difficult.
In contrast, our systematic initial condition treatment overcomes this difficulty.

Most surface-walking methods can only unsystematically find the connections between index-1 saddles and minima.
No serious effort is made to find the relations between saddles with different indices, on a family tree.
Exceptions are finding high-index saddles on the energy landscape without stating the pathway map and the bifurcation diagram \cite{doye2002saddle,li2001minimax, quapp2014locating}.
Here, we find that in our method, constructing the pathway map and the bifurcation diagram itself is important for saving the overall computation time leading to energy minima.

Our notion on the importance of stationary points for any general minimization problems can be supported by Ref.~\cite{mehta2011finding}, where a polynomial-like energy landscape is analyzed, in particular.
There, numerical polynomial homotopy continuation (NPHC) is made as the starting point, followed by solutions to reach the numerical result.
With the amount of classical B\'{e}zout bound (CBB) existing in this problem, the isolated stationary points of the energy landscape can be easily dealt with by using an efficient-path-tracker algorithm.
However, the NPHC method cannot be generalized beyond polynomial-like potentials to complex physical problems.
For example, to tackle the minimization problem of an energy functional, the variable space is broken down to an asymptotically large number of variables.
The problem is nonpolynomial type.
For the presented liquid-crystal problem in this Letter, each of the two unknown field variables, $Q_{11}(\bm{x})$ and $Q_{12}(\bm{x})$, is represented by an $N^2$-dimensional vector variable, where $N=100$ is the discretization grid number along each axis.
The number of the stationary points is estimated to be far less than the CBB ($3^{2N^2}$).
Since the NPHC method needs to track a CBB number of solutions, applying the NPHC method in this type of problems is expensive and unlikely.

{\it Supplemental Movies.}--- We include a movie, \texttt{downward.mp4}, to show the dynamic downward pathway sequence: $\mathrm{WORS} \rightarrow \mathrm{BD} \rightarrow \mathrm{BD}+\rightarrow I\rightarrow J-\rightarrow D$, and another movie, \texttt{upward.mp4}, to show the dynamic upward pathway sequence: $D \rightarrow J+\rightarrow H\rightarrow M+\rightarrow N+\rightarrow \mathrm{WORS}$ in Fig.~2(f) of the text.

\bibliography{WORSbib}